\documentclass[preprint2]{aastex}

\usepackage{natbib}
\usepackage{graphicx}

\begin{document}


\title{Identification of H$_2$CCC as a diffuse interstellar band carrier}

\shorttitle{H$_2$CCC as a DIB carrier}
\shortauthors{Maier et al.}

\author{J.\,P.\,Maier\altaffilmark{1}, G.\,A.\,H.\,Walker\altaffilmark{2,*}, D.\,A.\,Bohlender\altaffilmark{3,*}, F.\,J.\,Mazzotti\altaffilmark{1}, R.\,Raghunandan\altaffilmark{1}, J.\,Fulara\altaffilmark{1}, I.\,Garkusha\altaffilmark{1}, and A.\,Nagy\altaffilmark{1}}

\affil{$^1$ Department of Chemistry, University of Basel, Klingelbergstrasse 80, CH-4056 Basel, Switzerland; \email{j.p.maier@unibas.ch}}

\affil{$^2$ 1234 Hewlett Place, Victoria, BC V8S 4P7, Canada; \email{gordonwa@uvic.ca}}

\affil{$^3$ National Research Council of Canada, Herzberg Institute of Astrophysics,\\5071 West Saanich Road, Victoria, BC V9E 2E7, Canada; \email{david.bohlender@nrc-cnrc.gc.ca}}

\altaffiltext{*}{Based on observations obtained at the Canada$-$France$-$Hawaii Telescope (CFHT) which is operated by the National Research Council of Canada, the Institut National des Sciences de l'Univers of the Centre National de la Recherche Scientifique of France and the University of Hawaii, and at the Dominion Astrophysical Observatory, Herzberg Institute of Astrophysics, National Research Council of Canada.}

\begin{abstract}
We present strong evidence that  the broad, diffuse interstellar bands (DIBs) at 4881 and 5450\,\AA\/ are caused by the $B\,^1$B$_1$\,$\leftarrow$\,$X\,^1$A$_1$ transition of H$_2$CCC ($l$-C$_3$H$_2$). The large widths of the bands are due to the short lifetime of the $B\,^1$B$_1$ electronic state. The bands are predicted from absorption measurements in a neon matrix and observed by cavity ring-down in the gas phase and show exact matches to the profiles and wavelengths of the two broad DIBs. The strength of the 5450\,\AA\/ DIB leads to a $l$-C$_3$H$_2$ column density of $\sim$5$\times$10$^{14}$ cm$^{-2}$ towards HD\,183143 and $\sim$2$\times$10$^{14}$\,cm$^{-2}$ to HD\,206267. Despite similar values of $E$($B-V$), the 4881 and 5450\,\AA\/ DIBs in HD\,204827 are less than one third their strength in HD\,183143, while the column density of interstellar C$_3$ is unusually high for HD\,204827 but undetectable for HD\,183143. This can be understood if C$_3$ has been depleted by hydrogenation to species such as $l$-C$_3$H$_2$ towards HD\,183143. There are also three rotationally resolved sets of triplets of $l$-C$_3$H$_2$ in the 6150$-$6330\,\AA\/ region. Simulations, based on the derived spectroscopic constants and convolved with the expected instrumental and interstellar line broadening, show credible coincidences with sharp, weak DIBs for the two observable sets of triplets. The region of the third set is too obscured by the $\alpha$-band of telluric O$_2$.
\end{abstract}


\section{Introduction}
Diffuse absorption features were first noted in the spectra of `red' B stars some 90 years ago, but only recognized as arising from the interstellar medium in the 1930s. Initially, broad bands some tens of \AA\/ wide were identified, but with improved instrumentation many hundreds were subsequently found---mostly in the yellow/red with each being only a few \AA\/ wide. Although the strength of the diffuse interstellar bands (DIBs) is correlated roughly with colour excess, none shows the characteristics expected for absorption within the dust grains themselves. Rather, they are presumed to arise from gaseous molecules. Their characteristics have been summarised \citep{Snow:2006}. Several hypotheses have been put forward for their origin; most prominent being carbon chains and polycyclic hydrocarbon cations \citep{Herbig:2000}. Advances in laboratory techniques over the last decade have enabled the measurement of a number of such electronic spectra in the gas phase. These include some bare carbon chains \citep{Maier:2004}, those containing one hydrogen, nitrogen and their ions \citep{Motylewski:2000}, and a few aromatic cations \citep{Biennier:2003}. Individual coincidences of three very weak DIBs with the gas-phase absorptions of naphthalene cation \citep{Iglesias:2008} and, recently, for one band of diacetylene cation \citep{Krelowski:2010} have been reported.

A number of di- and triatomics are known to be present in the diffuse clouds: CH, CH$^+$, CN as well as H$_3^+$ and C$_3$ (see \citealp{Snow:2006} for references therein). In addition, mm-wave observations indicate the presence of more complex carbon containing molecules, e.g. $c$-C$_3$H$_2$ \citep{Cox:1988}, C$_4$H \citep{Liszt:2000}.

Recently, a match between a broad DIB, at 5450\,\AA\/ with $\sim$10\,\AA\/ FWHM, and the absorption of an unknown species produced in a plasma discharge of acetylene was reported \citep{Linnartz:2010}. The available experimental data pointed to a molecule containing only carbon and hydrogen. Here, we present evidence that the molecule responsible is propadienylidene, H$_2$CCC---linear three carbon atoms with two hydrogens off-axis on the terminal carbon; henceforth and usually designated as $l$-C$_3$H$_2$. The molecule has two other broad absorption bands at around 4890 and 5170\,\AA\/, close to known DIBs, as well as rotationally resolved bands at wavelengths above 6000\,\AA\/. Narrow DIBs of the expected strength are also found at the positions of two of the triplet systems predicted to be centred near 6159 and 6252\,\AA\/. The spectral region for the third system, weaker than the other two, is too confused by strong telluric oxygen features. All of this is evidence for an identification for a carrier molecule which gives rise to a series of DIBs.

The recognition that the broad band seen in the lab which coincided with the 5450\,\AA\/ DIB was due to absorption by $l$-C$_3$H$_2$ became evident from a number of previous studies, followed by the ones reported here. In its electronic ground state $X\,^1$A$_1$, $l$-C$_3$H$_2$ (C$_{\rm 2v}$ symmetry) has been well characterized by mm-wave spectroscopy in the laboratory \citep{Vrtilek:1990} and identified by this signature in dense interstellar clouds \citep{Cernicharo:1991}, and in the diffuse medium \citep{Cernicharo:1999}. It is an isomer of the ubiquitous cyclic C$_3$H$_2$ \citep{Madden:1989}. The electronic absorption spectrum of $l$-C$_3$H$_2$ was first observed in a neon matrix at 6\,K \citep{Hodges:2000}. In the visible region it shows two electronic systems; a weak dipole forbidden $A\,^1$A$_2$\,$\leftarrow$\,$X\,^1$A$_1$ transition which becomes partly allowed by vibronic effects with the near lying $B\,^1$B$_1$ electronic state. Then it shows an intense $B\,^1$B$_1$\,$\leftarrow$\,$X\,^1$A$_1$ transition in the 4000$-$5500\,\AA\/ range. The spectrum has now been re-recorded following mass-selected deposition of C$_3$H$_2^+$ cations followed by neutralization; the earlier measured spectrum was obtained by photolysis of a chemical precursor in the matrix. The spectra are by and large identical, except that the present recording (Fig.\,1) shows better resolution of the bands and there are wavelength errors for some of the reported absorptions. Due to matrix-gas shifts, a direct comparison with DIB data is inconclusive.

Thus a goal a number of years ago was to observe the $B\,^1$B$_1$\,$\leftarrow$\,$X\,^1$A$_1$ system of $l$-C$_3$H$_2$ in the gas phase. Three techniques were employed: absorption measurements in a supersonic jet expansion through which a discharge runs and detection with cavity ring-down (CRD) either with pulsed- or continuous lasers, as well as resonance enhanced multi-photon ionisation. The experiments were unsuccessful \citep{Achkasova:2006}, from which it was concluded that the $B\,^1$B$_1$ excited electronic state has a lifetime less than a picosecond, causing individual line broadening of over 10\,cm$^{-1}$. However, at the same time, three rotationally resolved bands lying in the 6000$-$6600\,\AA\/ region were observed \citep{Achkasova:2006,Birza:2005}. In the latter two articles, the rotational analyses proved that these belong to $l$-C$_3$H$_2$, corresponding to transitions from the $v$=0 level of the ground $X\,^1$A$_1$ state to vibronic levels in the $A\,^1$A$_2$ state, lying near the origin of the $B\,^1$B$_1$ electronic state. The bands in the dipole forbidden $A\,^1$A$_2$\,$\leftarrow$\,$X\,^1$A$_1$ transition gain intensity by vibronic mixing with the $B\,^1$B$_1$ levels. The rotational analysis is unambiguous. These bands, lying around 6159, 6252 and 6319\,\AA\/ in the gas phase show a complex structure characteristic of an asymmetric top.

\section{Laboratory Observations}

\subsection{Techniques}
The two laboratory techniques used were absorption measurements in a neon matrix and CRD in the gas phase. For the former C$_3$H$_2^+$ was produced from allene in an electron impact ion source. The mass-selected $m$/$e$=38 ions were then co-deposited with excess of neon onto a sapphire surface at 6\,K to form the matrix \citep{Freivogel:1994}. This took a few hours using nA ion currents to produce a matrix of 150\,$\mu$m thickness, spread over a few cm$^2$. The ions were subsequently neutralized by UV irradiation of the matrix, releasing electrons from the counter ions present. The absorption spectra were measured by passing light through the thin side of the matrix over a path length of 2\,cm, dispersion of the transmitted light by a grating and detection with a CCD camera. Another absorption spectrum was obtained using allene as precursor in the matrix followed by xenon lamp VUV (1470\,\AA\/) photolysis. The mass-selected measurements on $l$-C$_3$D$_2$ were achieved with dideutero-acetylene in the ion source.

In the gas phase $l$-C$_3$H$_2$ was produced in a supersonic free-jet expansion discharge. As precursor, 0.5$-$1\,$\%$ allene in argon, or acetylene and diacetylene in helium, was used. The CRD set up is as described \citep{Linnartz:1998}. The discharge was operated at 10\,Hz while the laser was fired every 20\,Hz. The CRD signal, which is not background-free, could then be corrected for baseline. It was found that the strongest signals were observed with allene, but at the same time the vibrational and rotational temperatures attained were a bit higher than with diacetylene. Furthermore, allene produced other species causing broad absorptions in the range of interest (e.g. $l$-C$_3$H). Dideutero-acetylene in helium was used in the discharge expansion for the measurements on $l$-C$_3$D$_2$.

\subsection{Laboratory Results}
We focused on whether the 5450\,\AA\/ absorption band coinciding with the DIB at this wavelength \citep{Linnartz:2010} could be the $B\,^1$B$_1$\,$\leftarrow$\,$X\,^1$A$_1$ transition of $l$-C$_3$H$_2$, i.e. corresponding to the most intense band in the 6\,K neon matrix spectrum at 5417\,\AA\/ (Fig.\,1). The middle trace was obtained using C$_3$H$_2^+$ mass selection followed by neutralization. That this is the absorption of linear C$_3$H$_2$ has been proven \citep{Hodges:2000}. Our spectrum is the same, but with a better resolution; for example the site-structured bands show two components and the correct calibration of the 5445/5417\,\AA\/ band is given in Table\,1. The spectrum below 4700\,\AA\/ is not shown, because overlapping bands are present there due to $l$-C$_3$H \citep{Ding:2001}, produced as a result of H loss during the deposition of C$_3$H$_2^+$.

The top trace is the spectrum obtained via a traditional matrix approach: allene was embedded in neon and following photolysis using a low-pressure xenon lamp the system appears. This spectrum is identical to the one published \citep{Hodges:2000}, though produced by a different chemical scheme and irradiation. It is included because the intensity ratio of the respective bands is more reliable than in the mass-selected spectrum where the absorption involves much smaller concentrations and thus background correction is not so reliable. The doublet at 5445/5417\,\AA\/ is not resolved but is indicated from the asymmetry of the band. The structure of the 5417\,\AA\/ band is due to matrix site effect, because the same asymmetric pattern is seen for the other bands (5143, 4856 and 4633\,\AA\/).

Bottom trace in Fig.\,1 shows the spectrum measured after deposition and neutralization of C$_3$D$_2^+$ ions into a neon matrix. The site structure is not present for the prominent band at 5412\,\AA\/; such an effect is often seen on isotopic substitution in matrices. The band at 5412\,\AA\/ is shifted by 17\,cm$^{-1}$ towards higher energy in comparison to the 5417\,\AA\/ band of $l$-C$_3$H$_2$. 

The assignment of the bands to vibrational excitation in the upper electronic state of $l$-C$_3$H$_2$ is established \citep{Hodges:2000}. All the transitions originate from the $v$=0 level of the $X\,^1$A$_1$ ground state because the temperature of the neon matrix is 6\,K. The strongest 5417\,\AA\/ band corresponds to the excitation of the totally symmetric $\nu_2$ (C=C stretching) mode in the $B\,^1$B$_1$ state (2$_0^1$ transition). Other strong bands in the spectrum of C$_3$H$_2$ are due to its overtone (2$_0^2$ at 4856\,\AA\/) and combination with the $\nu_4$ mode (2$_0^1$\,4$_0^1$ at 5143\,\AA\/) as indicated in Fig.\,1. The origin band of the $B\,^1$B$_1$\,$\leftarrow$\,$X\,^1$A$_1$ transition is weaker and is influenced by vibronic interaction of $B\,^1$B$_1$ with the nearby lying $A\,^1$A$_2$ state. It is located somewhere within the group of weak bands above 6000\,\AA\/ (between a and c in Fig.\,1). This has been considered theoretically \citep{Hodges:2000} and also discussed in the analysis of the rotationally resolved spectra in the gas phase in this region \citep{Achkasova:2006}.

CRD is a sensitive detection method; however, the lack of mass selection is a drawback. Therefore, the matrix spectra of the mass-selected C$_3$H$_2$ and C$_3$D$_2$ serve for the identification of the bands in the gas-phase CRD experiments. Though the wavelength shift going from neon matrix to the gas phase can be 10$-$50\,\AA\/ for electronic transitions, the separation of vibrational peaks rarely differ by more than a few \AA\/ within one state \citep{Jacox:1994}. Thus, by knowing the wavelength of the 2$_0^1$ transition in the gas phase, the positions of other bands (e.g. 2$_0^2$ or 2$_0^1$\,4$_0^1$) can be predicted from the separation of the peaks in the neon spectrum (Fig.\,1). However, this procedure cannot be applied for the 2$_0^1$ band of $l$-C$_3$H$_2$ itself because the separation from the origin to the 2$_0^1$ peak is not known.

The aim was therefore to detect the gas-phase band of $l$-C$_3$H$_2$ that corresponds to the 2$_0^1$ transition in the neon matrix at 5417\,\AA\/. The known rotationally resolved bands of $l$-C$_3$H$_2$ above 6000\,\AA\/ were first detected using acetylene, diacetylene or allene seeded in argon in the discharge. The strongest signal was observed with allene. The spectrum was then measured in the 5420$\pm$50\,\AA\/ range as indicated by the 2$_0^1$ absorption in neon. Broad absorptions were detected with all the precursors and under the same discharge conditions as for the rotationally resolved $\lambda>$6000\,\AA\/ bands of $l$-C$_3$H$_2$. The spectrum obtained with allene precursor is shown in Fig.\,2\,a. It consists of two broad absorptions, whose relative intensities vary with the precursor proving that they originate from two different molecules. The intensity of the band at lower wavelength ($\sim$5450\,\AA\/) correlates with the identified absorptions of $l$-C$_3$H$_2$ in the range above 6000\,\AA\/, but the one around 5480\,\AA\/ does not.

The absorptions in Fig.\,2\,a have been deconvoluted using the two Gaussians shown. The central wavelength of the first peak is $\sim$5450\,\AA\/ (FWHM $\sim$13$-$17\,\AA\/), whereas the other is centered at 5473$-$5476\,\AA\/ (FWHM $\sim$32$-$42\,\AA\/). The prominent C$_2$ absorption lines have been partially removed from the spectrum by taking a wavelength window of ten points, finding the minimum and then repeating the procedure at the following position of the spectrum. By this means the broad underlying absorption is retained, and fitted by a Gaussian. Fig.\,2\,a shows the superposition of the two Gaussians and the quality of the fit.

Gas-phase measurements on $l$-C$_3$D$_2$ were also carried out. In the spectral window of interest only one broad band (central wavelength at 5458(3)\,\AA\/ and FWHM $\sim$19$-$24\,\AA\/) has been detected using dideutero-acetylene as the precursor (Fig.\,2\,b). The FWHM does not differ much from that of the 5450\,\AA\/ band. The position of this band is shifted $\sim$27\,cm$^{-1}$ to the lower energy relative to the 5450\,\AA\/ band, similar to the neon matrix change (17\,cm$^{-1}$).

Comparison of the neon matrix spectra of $l$-C$_3$H$_2$ and $l$-C$_3$D$_2$ (Fig.\,1) with the CRD spectra (Fig.\,2) leads to the conclusion that the 5417\,\AA\/ band in the neon matrix and that at 5450\,\AA\/ in the gas phase originate from the same molecule, namely $l$-C$_3$H$_2$. Also the fact that the intensity of the band at 5450\,\AA\/ observed in the gas phase, correlates with the intensity of other, already identified bands of $l$-C$_3$H$_2$ in the range above 6000\,\AA\/, substantiates this conclusion.

From the position of the 2$_0^1$ band of $l$-C$_3$H$_2$ in the gas phase (5450\,\AA\/) the wavelengths of the 2$_0^2$ and 2$_0^1$\,4$_0^1$ transitions in the gas phase can be predicted using the separation of the peaks in the neon spectrum. For this the most intense peaks are taken (5417, 5143 and 4856\,\AA\/), because the lowest energy site is not resolved (though its presence is indicated by the asymmetry of the peak). The difference between the 2$_0^1$ and 2$_0^2$ band maxima is 2130\,cm$^{-1}$ and this leads to a gas phase prediction of 4883\,\AA\/ for the 2$_0^2$ transition; the uncertainty is difficult to guess because of broadening mechanisms in the matrix. The 2$_0^1$\,4$_0^1$ band is located 984\,cm$^{-1}$ above the   transition in the neon matrix and in the gas phase it is then in the 5173\,\AA\/ wavelength region.

In the region around 4883\,\AA\/ predicted for the 2$_0^2$ band, broad absorptions are detected (Fig.\,2\,c) with narrower components imposed on them. The four bands at 4913, 4870, 4856 and 4844\,\AA\/ belong to $l$-C$_3$H because they have been identified in earlier mass-selected studies in the gas phase \citep{Ding:2001}. After subtraction of the $l$-C$_3$H absorptions (the 4 Gaussians shown in Fig.\,2\,c) from the laboratory recording, the residual broad band is obtained (Fig.\,2\,d). The Gaussian fit is centered at 4887(3)\,\AA\/ with FWHM of $\sim$25\,\AA\/. Its position matches the 4883\,\AA\/ wavelength predicted from the neon matrix spectrum of $l$-C$_3$H$_2$. The FWHM of this band is somewhat larger than that of the 2$_0^1$ band at 5450\,\AA\/ probably due to a shorter lifetime in the $v_2$=2 vibrational state of $B\,^1$B$_1$. These arguments, together with the fact that the 5450 and 4887\,\AA\/ bands were observed under the same experimental conditions as the rotationally resolved transitions of $l$-C$_3$H$_2$ above 6000\,\AA\/ provide the evidence that these absorptions originate from the $l$-C$_3$H$_2$ molecule.

The detection of the band around 5173\,\AA\/, corresponding to the 2$_0^1$\,4$_0^1$ transition of $l$-C$_3$H$_2$, was not successful. This is because this transition is a factor of three weaker than 2$_0^1$ and its intensity is below the detection limit of the CRD set-up.

It was already established \citep{Linnartz:2010} that the 5450\,\AA\/ gas-phase absorption coincides with a DIB at this wavelength. This band is now assigned to the 2$_0^1$ $B\,^1$B$_1$\,$\leftarrow$\,$X\,^1$A$_1$ transition of $l$-C$_3$H$_2$. A DIB corresponding to the 2$_0^2$ band in the gas-phase observations is present (see Section\,4).

\section{Simulation of Band Profiles}

\subsection{Bands Above 6000\,\AA\/}
Three systems in the 6150$-$6330\,\AA\/ region have been subject to two earlier high-resolution, gas-phase studies \citep{Achkasova:2006,Birza:2005}. The rotational analysis is unambiguous and proves that the absorptions belong to $l$-C$_3$H$_2$. 

$l$-C$_3$H$_2$ has a near prolate top structure with $C_{2v}$ symmetry. The rotational constants in the $X\,^1$A$_1$ ground state are $A$=9.6451, $B$=0.3533 and $C$=0.3404\,cm$^{-1}$. The $K$=1\,$-$\,$K$=0 and $K$=2\,$-$\,$K$=0 separations are $\sim$10 and $\sim$40\,cm$^{-1}$ respectively. Thus at very low temperatures only the $K$=0,1 levels are populated. According to the selection rules only transitions with $\Delta K$=$\pm$1 are observed.

In Fig.\,3, left, are shown the simulations of the bands of $l$-C$_3$H$_2$ above 6000\,\AA\/, using the derived spectroscopic constants and a Lorentzian profile of 0.15\,\AA\/ width, corresponding to the resolving power 40,000 used in astronomical measurements, for each individual line. The pattern is shown for temperatures of 5, 20 and 60\,K. Three components, $K$=2$\leftarrow$1, $K$=1$\leftarrow$0 and $K$=0$\leftarrow$1 would be observed. At 5\,K the population becomes concentrated in the $K$=0 level of the $X\,^1$A$_1$ electronic ground state and thus the middle peak dominates. As the temperature rises the outer two peaks gain intensity because the population of the $K$=1 level increases.

Though the resolving power in many astronomical measurements is around 40,000, observations such as toward HD\,206267 sample several clouds as the profiles of K and Ca lines show \citep{Pan:2004} and this leads to broadenings of 10$-$20\,km\,s$^{-1}$, i.e. 0.2$-$0.4\,\AA\/ in this wavelength region. Thus the absorption pattern of $l$-C$_3$H$_2$ in such clouds would look like that shown in Fig.\,3, right, where a 0.4\,\AA\/ width (Lorentzian) is used in the simulation.

\subsection{5450\,\AA\/ Band}
The 5450 and 4887\,\AA\/ absorption bands of $l$-C$_3$H$_2$ are broad, without any rotational structure. This arises due to the short lifetime of the $B\,^1$B$_1$ electronic state as a result of fast intramolecular relaxation process to the nearby lying $A\,^1$A$_2$ state. The profiles of the broad 5450 and 4887\,\AA\/ bands can be predicted by simulating the 2$_0^1$ and 2$_0^2$ $B\,^1$B$_1$\,$\leftarrow$\,$X\,^1$A$_1$ transitions with the known rotational constants for the bands above 6000\,\AA\/ \citep{Achkasova:2006}. Due to the broadness of the features, the precise value of the rotational constants has little influence on the overall contour. The bottom trace in Fig.\,4 shows the case with a 1\,\AA\/ linewidth corresponding to 1\,ps lifetime and temperature of 10\,K. Three components in the profile of this band would still be seen.

The best fit to the observed 5450\,\AA\/ profile was obtained with a 10\,\AA\/ linewidth and 60\,K temperature (top trace in Fig.\,4). The temperature of 60\,K is chosen because the rotationally resolved bands of $l$-C$_3$H$_2$ measured above 6000\,\AA\/ fit to 50$-$70\,K. The linewidth of 10\,\AA\/ corresponds to a 100\,fs lifetime, which is often encountered in the excited electronic states of polyatomic molecules subject to conical intersections. In fact in the case of $l$-C$_3$H$_2$ the next higher lying $C\,^1$A$_1$ electronic state, with onset near 2500\,\AA\/, has been shown experimentally to have a lifetime of 70\,fs \citep{Noller:2009}, so the inferred value of 100\,fs is not unreasonable.

\section{Astronomical Observations}

\subsection{Broad Bands}
Broad absorptions are observed in the lab at 4887 and 5450\,\AA\/ in the gas phase and one predicted in the 5165$-$5185\,\AA\/ range but expected to be only one half to one third the strength of the 5450\,\AA\/ absorption. The absorption spectrum in Fig.\,1 is a linear measurement and thus predicts that any corresponding DIBs should have comparable EWs, and FWHM. There is a strong, broad DIB at 4881\,\AA\/ with twice the FWHM ($\sim$25\AA\/), but similar central depth to the 5450\,\AA\/ DIB. It is listed in many DIB compilations \citep{Jenniskens:1994,Hobbs:2008,Herbig:1995} and was recorded anew as part of this study. Unfortunately, stellar H$\beta$ (4861\,\AA\/) contaminates its short-wavelength wing.   

Spectra were acquired in the regions of these DIBs at the Dominion Astrophysical Observatory with the 1.2 and 1.8\,m telescopes at a resolution of $R$=18,000 in early 2010. The reddened target stars are listed in Table\,2 and the spectra displayed in Figs.\,5 and 6 for the 4881 and 5450\,\AA\/ regions, respectively, together with those of unreddened stars. The spectra for individual stars were aligned to the mean rest wavelength of the interstellar K\,I line in each case.

\begin{figure}
 \plotone{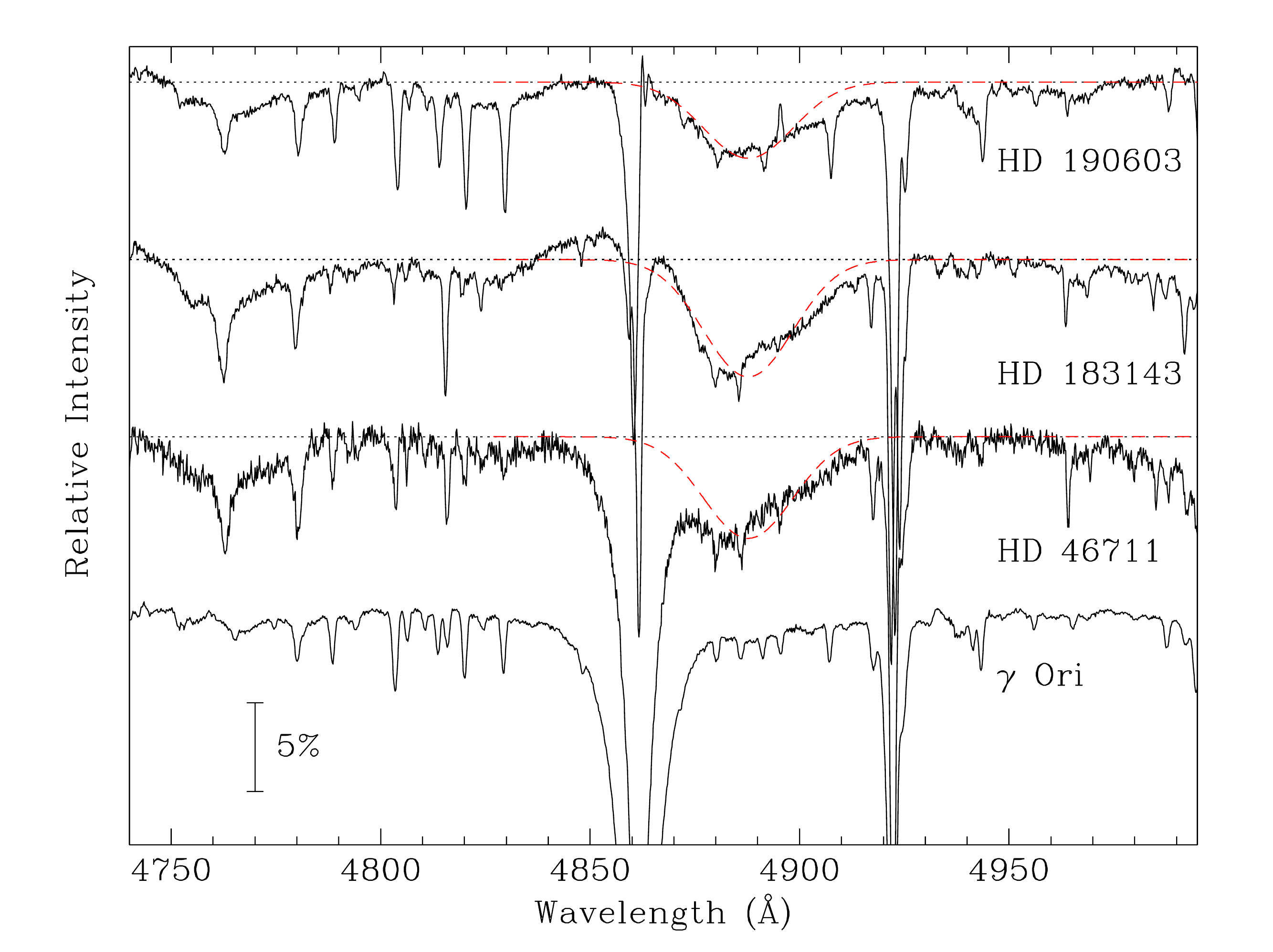}

\caption{The 4881\,\AA\/ DIB, already known from the literature, in spectra of HD\,46711, 190603, 183143 and $\gamma$\,Ori (unreddened standard), is compared to a Gaussian fit (FWHM 25\,\AA) to the laboratory absorption spectrum of $l$-C$_3$H$_2$ centred at 4887\,\AA\/  (dashed, red line).  The short wavelength side of the DIB profile is distorted to varying degrees by H$\beta$ at 4861\,\AA. \label{4880} }
\end{figure}

The coincidence of the lab and DIB profiles in the case of the 5450\,\AA\/ band has already been discussed \citep{Linnartz:2010}. The Gaussian (25\,\AA\/ FWHM) obtained from the CRD lab measurement of the 4887\,\AA\/ absorption (Section\,2.2) is superimposed on the DIB spectra in Fig.\,\ref{4880}. The agreement is again excellent. The wings of the strong stellar H$\beta$ line at 4861\,\AA\/ makes that part of the DIB hard to define for HD\,46711. For the supergiant, HD\,183143, there might also be a broad emission component.

The case for a FWHM$\sim$20\,\AA\/ DIB in the 5165$-$5185\,\AA\/ range is complicated by the presence of a previously known, and well defined, broad DIB centred at 5160\,\AA\/. Unfortunately, the presence of strong stellar lines near 5173\,\AA\/ in all of our target stars made it impossible to say anything about an interstellar absorption feature in that region.

\subsection{Narrow Triplets Above 6000\,\AA\/}
The triplets offer quite a different challenge from the broad bands when it comes to unambiguous detection. An exact wavelength match was less critical for the broad bands. There are many fewer broad DIBs known than sharp ones making chance coincidence less likely. Also, the broad DIBs can be matched in width without significant distortion from spectral resolution and blurring by the velocity dispersion in the interstellar clouds. For the 4881 and 5450\,\AA\/ regions there is little contamination by telluric lines and stellar lines are easily distinguished. By contrast, the spectra of the late-type B giants are full of stellar lines which are hard to distinguish from sharp DIBs. The highest density of sharp DIBs is in the 6000\,\AA\/ region, one every few \AA\/, which makes the chance of coincidence likely. There are also many telluric lines. It should also be noted that the sharp DIBs of interest have central intensities of $<$1\,\%.

As a result of these obstacles, we were restricted to reddened early B or O stars and spectra with a resolution $>$40,000 and S/N of several thousand per \AA\/. We drew on CFHT archival spectra and others were especially acquired of HD\,179406 with the CFHT ESPaDONs spectrograph in June 2010. The stars are listed in Table\,3 where `Gec' corresponds to Gecko with a median resolution of 120,000 and `Esp' is ESPaDONs with $R\sim$80,000. Both are bench spectrographs at the CFHT. It should be emphasized that the archival spectra were optimized for other programs. Low order polynomials have been applied to remove residual undulations in the final spectra.

\begin{figure}
 \plotone{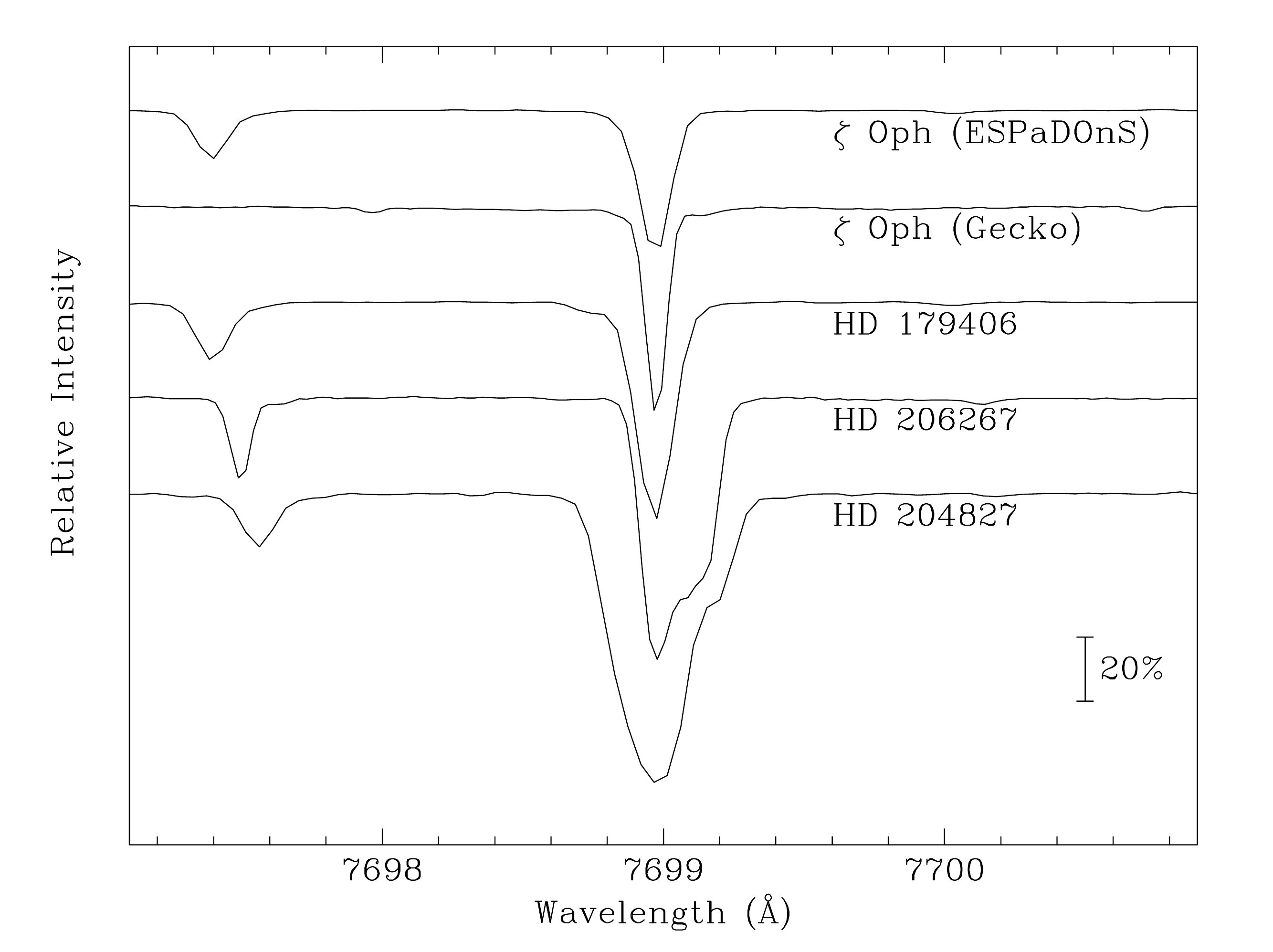}

\caption{Interstellar K\,I line (7699\,\AA\/) profiles for the reddened target stars aligned on their strongest interstellar component. The K\,I radial velocities actually adopted are given in Table\,3.\label{KI} }
\end{figure}

All of the target stars are seen through multiple interstellar clouds with some differential velocity ranges $>$10\,km\,s$^{-1}$ as can be seen in Fig.\,\ref{KI} where the interstellar K\,I line profiles are plotted and aligned on their strongest components. This alignment was applied to the individual stellar spectra and the corresponding interstellar velocities are listed in Table\,3.

\begin{figure}
 \plotone{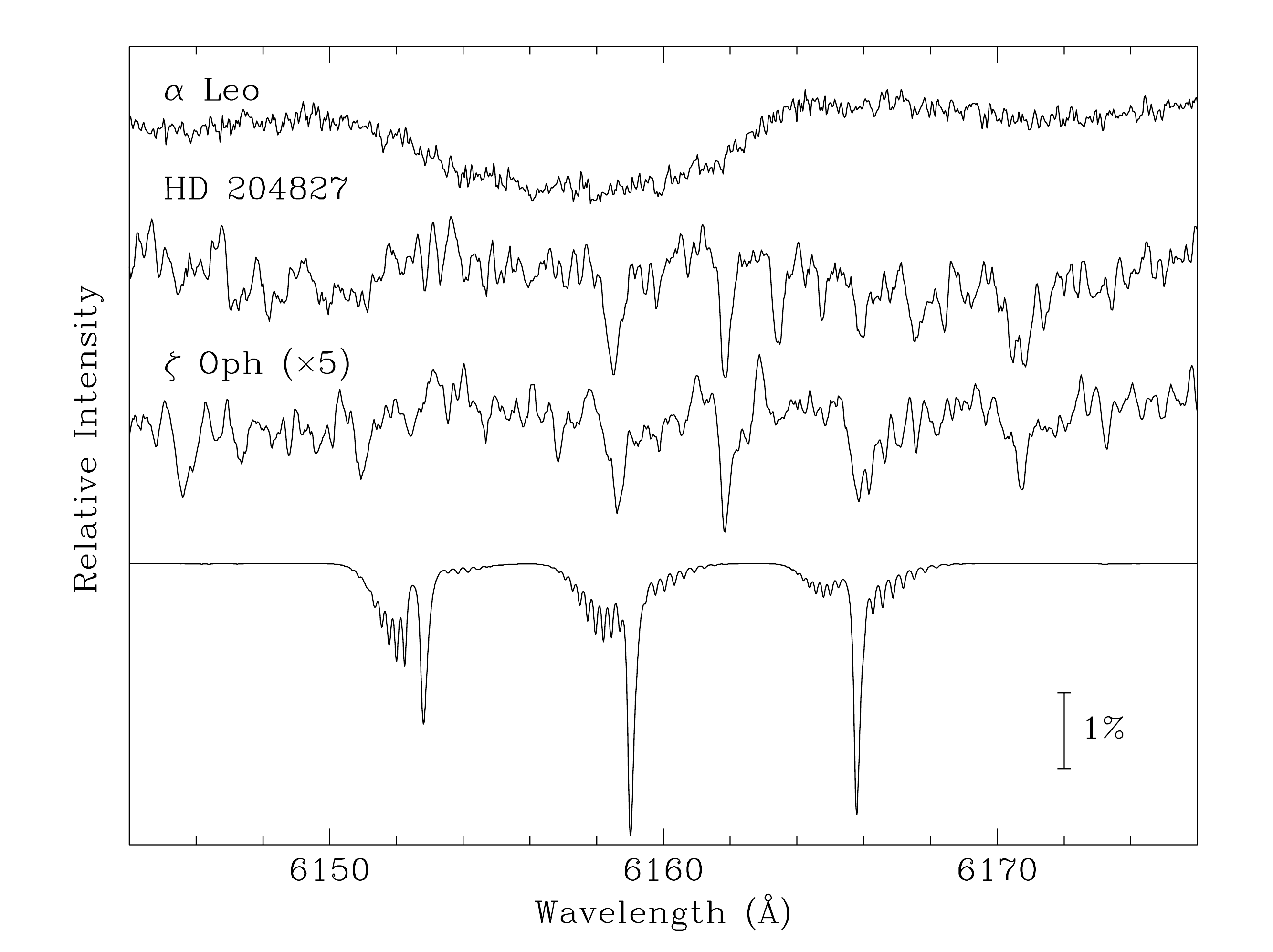}

\caption{Spectra of HD\,204827, $\zeta$\,Oph, the unreddened standard, $\alpha$\,Leo and model line profiles for $l$-C$_3$H$_2$ at 10\,K and 0.15\,\AA\/ FWHM. The spectrum for $\zeta$\,Oph has been expanded by a factor of 5 relative to the others. The broad feature centred at 6159\,\AA\/ in $\alpha$\,Leo is a rotationally broadened stellar line. Narrow DIB lie within the 6159 and 6166\,\AA\/ model profiles for the two reddened stars.\label{6150} }
\end{figure}

\begin{figure}
 \plotone{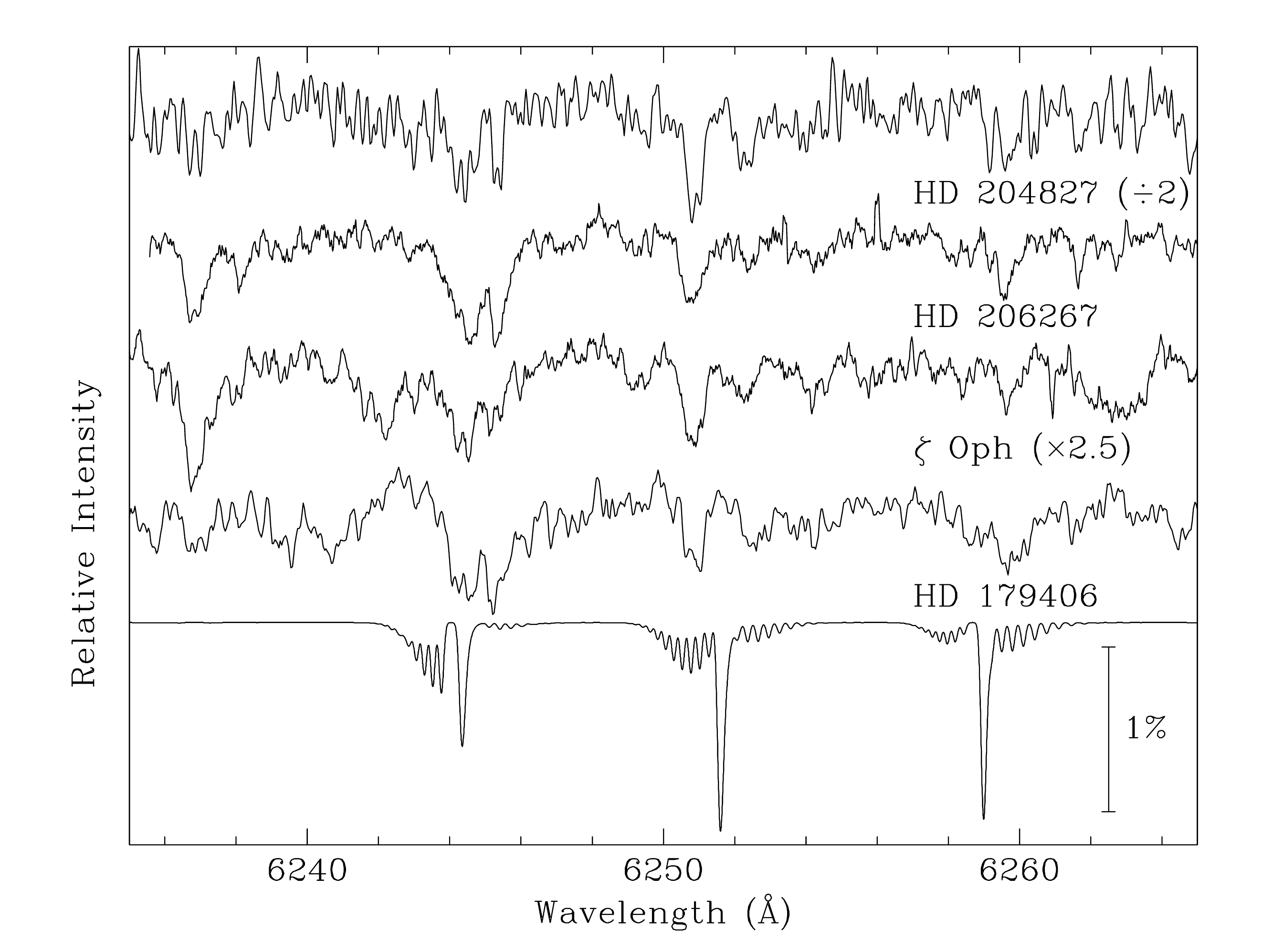}

\caption{Spectra of HD\,204827, HD\,206267, $\zeta$\,Oph, and HD\,179406 together with model line profiles for $l$-C$_3$H$_2$ at 10\,K and 0.15\,\AA\/ FWHM. The spectrum for $\zeta$\,Oph has been expanded by a factor of 2.5 while that for HD\,204827 has been halved. There are DIB consistently within the limits of the model line profiles.\label{6250} }
\end{figure}

Figs.\,\ref{6150} and \ref{6250} display the spectra of the reddened stars for two sets of the predicted $l$-C$_{3}$H$_{2}$ triplets together with model absorption profiles for 10\,K and a 0.15\,\AA\/ FWHM, the latter to account for instrumental and interstellar broadening. There are clearly DIBs within the limits of the model profiles at 6159, 6166, 6244, 6251 and 6259\,\AA\/. The line predicted at 6152\,\AA\/ was expected to be significantly weaker than those at 6159 and 6166\,\AA\/ so the absence of a DIB at that wavelength is not necessarily a contradiction. The DIBs at 6244 and 6251\,\AA\/ appear to be double with the stronger component corresponding quite closely to the predicted absorption maximum. The comparison between the lab and interstellar wavelengths together with the EW of the individual astronomical DIB for each of the target stars are given in Table 4.

For the triplets predicted near 6320\,\AA\/, the region is heavily obscured by the telluric $\alpha$-band of O$_2$ which, coupled with velocity corrections to the barycentre and alignment to the interstellar K\,I velocities, makes it impossible to detect any DIB which might coincide with the predictions.

In none of the coincidences do the predicted profiles correspond well to those of the DIBs, suggesting that the intrinsic broadening is probably greater than the 0.15\,\AA\/ assumed for the model. While the coincidences between DIBs and the model profiles are not unequivocal given the high density of sharp DIBs in this spectral region, their absence would have cast some doubt on the case $l$-C$_3$H$_2$ as a DIB carrier.

\section{Discussion and Conclusions}

The choice of a rotational temperature for $l$-C$_3$H$_2$ in diffuse clouds has a consequence for the comparison with the laboratory data, as the simulations in Figs.\,3 and 4 show. $l$-C$_3$H$_2$ has a large dipole moment (around 4\,D) and may cool efficiently, in much the same way as do the diatomics CH, CH$^+$ and CN, which attain 2.7\,K in the diffuse medium. On the other hand, the temperature inferred for other polar polyatomics detected in the diffuse medium are often higher, for example around 50\,K for ammonia by mm-wave spectroscopy \citep{Liszt:2006} and H$_3^+$ by IR measurements \citep{McCall:2002}. Due to the broadness of the 5450\,\AA\/ DIB the interstellar temperature of $l$-C$_3$H$_2$ can not be clearly determined; anything in the range 10$-$60\,K would fit as the two simulated profiles with 10\,\AA\/ linewidth in Fig.\,4 show.

As a follow up to the identification of $l$-C$_3$H$_2$ as a DIB carrier, one can make an estimate for the column density on the basis of the broad 5450\,\AA\/ DIB. There are two calculations \citep{Mebel:1998} of the overall oscillator strength, $f$, of the $B\,^1$B$_1$\,$\leftarrow$\,$X\,^1$A$_1$ electronic transition, 0.016 and 0.009. The former value was obtained with a higher level method and is used for the evaluation.

The $f$ value for the 2$_0^1$ transition (5450\,\AA\/ band) within this system is estimated from the calculation of the Franck$-$Condon factors; these are shown in Fig.\,5 of \citep{Mebel:1998}. This gives $f$(2$_0^1$)/$f_{\rm total}$(=0.016)\,=\,0.2. Using the expression:

\[
N(\textrm{cm}^{-2}) = 1.13 \times10^{20} \frac{\textrm{EW(\AA\/)}}{\Lambda\textrm{(\AA\/)}^2\times f}
\] \\
yields for $\Lambda$=5450\,\AA\/, $f$(2$_0^1$)=0.003, EW=0.4\,\AA\/ towards HD\,183143: $N$($l$-C$_3$H$_2$)$\approx$5$\times$10$^{14}$\,cm$^{-2}$.

Towards HD\,183143 the column density of H$_3^+$ has been determined to be 10$^{14}$\,cm$^{-2}$ \citep{McCall:2002} and $N$(C$_2$)$\approx$10$^{14}$\,cm$^{-2}$ \citep{Oka:2003}. Thus the $N$ value estimated for $l$-C$_3$H$_2$ is not unreasonable.

An obvious question is the role of $c$-C$_3$H$_2$. Radioastronomy has established that this cyclic isomer is ubiquitous in dense and diffuse interstellar media. But $c$-C$_3$H$_2$ does not have an allowed electronic transition in the DIB region. Its absorption has been observed in the laboratory, in a neon matrix with onset around 2700\,\AA\/ \citep{Seburg:1997}. Apart from $l$-C$_3$H$_2$, also $l$-C$_3$H has many absorption bands in the DIB region \citep{Ding:2001}. Some of the peaks are actually observed in the CRD spectra measured, nearby to the broad bands of $l$-C$_3$H$_2$. However, as has already been pointed out, due to the small oscillator strength of such individual transitions the DIB absorptions would have EW of m\AA\/ or less. Nevertheless, chemically the presence of $l$-C$_3$H in addition to the here proven $l$-C$_3$H$_2$ could be investigated with high S/N DIB measurements because the gas-phase data are available. Another related aspect concerns an earlier discussion on the possible presence of $l$-C$_3$H$_2^{-}$ in the diffuse medium \citep{Guethe:2001}. It was pointed out that $l$-C$_3$H$_2^{-}$ formation could be significant via associate electron attachment to $l$-C$_3$H$_2$.

As the 4881 and 5450\,\AA\/ DIBs correspond to the two broad absorptions of $l$-C$_3$H$_2$, a further aspect has to be considered, namely the estimate of the expected EW for the narrow b,\,c bands around 6251\,\AA\/ and 6159\,\AA\/ with the triplet structure in the DIB observations as discussed in Section\,4.2. The matrix absorption spectrum (Fig.\,1 and Table\,1) indicates from the peak areas that the b band should have $\sim$1/10 intensity of the 5450\,\AA\/ band. The intensity in the gas phase is distributed between the three $K$ components (Fig.\,3). It has to be decided what is the appropriate rotational temperature. As is seen in Fig.\,3, at 5\,K the middle peak, $K$=1$\leftarrow$0 component, will be the most intense; at 20\,K the outer two ($K$=2$\leftarrow$1 and 0$\leftarrow$1) already become stronger. This will thus influence a comparison of the DIB EW with laboratory data.

The column density of $l$-C$_3$H$_2$, $\sim$5$\times$10$^{14}$\,cm$^{-2}$---inferred in the previous section, was for the 5450\,\AA\/ DIB towards HD\,183143. For the observation towards HD\,206267 (Section\,4.1) the EW=0.13\,\AA\/ is lower and hence $N$($l$-C$_3$H$_2$)$\approx$2$\times$10$^{14}$\,cm$^{-2}$. We can now use this value to estimate the EW of the 6250\,\AA\/ band system (Fig.\,3). The oscillator strength of b is taken as $f$=0.0003, guessed by comparison with the corresponding absorption bands in the neon matrix. An uncertainty arises because of variation in phonon broadening and site structure (Fig.\,1). Then at 6250\,\AA\/ this leads to an EW of around 10\,m\AA\/. Thus below 5\,K only the $K$=1$\leftarrow$0 peak would be seen and have EW of this order of magnitude. At a higher temperature, 10$-$30\,K, this would be distributed between the three components $K$=2$\leftarrow$1, $K$=1$\leftarrow$0, $K$=0$\leftarrow$1, each with EW of a few m\AA\/.

An interesting inference concerns the anti-correlation with C$_3$ abundance in diffuse clouds. From our measurements of the 4881 and 5450\,\AA\/ DIBs in four stars covering a large range in reddening and supplemented by measurements of HD\,204827 from \citet{Hobbs:2008}, we find that these two DIBs correlate much more closely with each other than with $E$($B-V$). This implies a common carrier. \citet{Hobbs:2008} specifically observed HD\,204827 because of its exceptionally high column density of interstellar C$_3$. They could detect none towards HD\,183143. By contrast, from their spectra, the 4881 and 5450\,\AA\/ DIBs in HD\,204827 are less than one third their strength in HD\,183143. There is a similar, negative trend between the line of sight abundance of interstellar C$_3$ \citep{Oka:2003} and the strengths of the 4881 and 5450\,\AA\/ DIBs for the stars in our sample. The inference seems clear, hydrogenation of C$_3$ to species such as $l$-C$_3$H$_2$ leads to depletion of C$_3$ with a consequent enhancement of the 4881 and 5450\,\AA\/ DIBs. This is in contrast to the family of enhanced DIBs found by \citet{Thorburn:2003} for HD\,204827 which correlates closely with $N$(C$_2$).

The present study provides laboratory and astronomical evidence for $l$-C$_3$H$_2$ in the diffuse interstellar clouds and its identification as a DIB carrier.

\acknowledgments
This work was supported by the Swiss National Science Foundation (project 200020-124349/1). GAHW and DAB thank the Director of the CFHT for one hour of discretionary time to observe HD\,179406.

\begin{figure}
\plotone{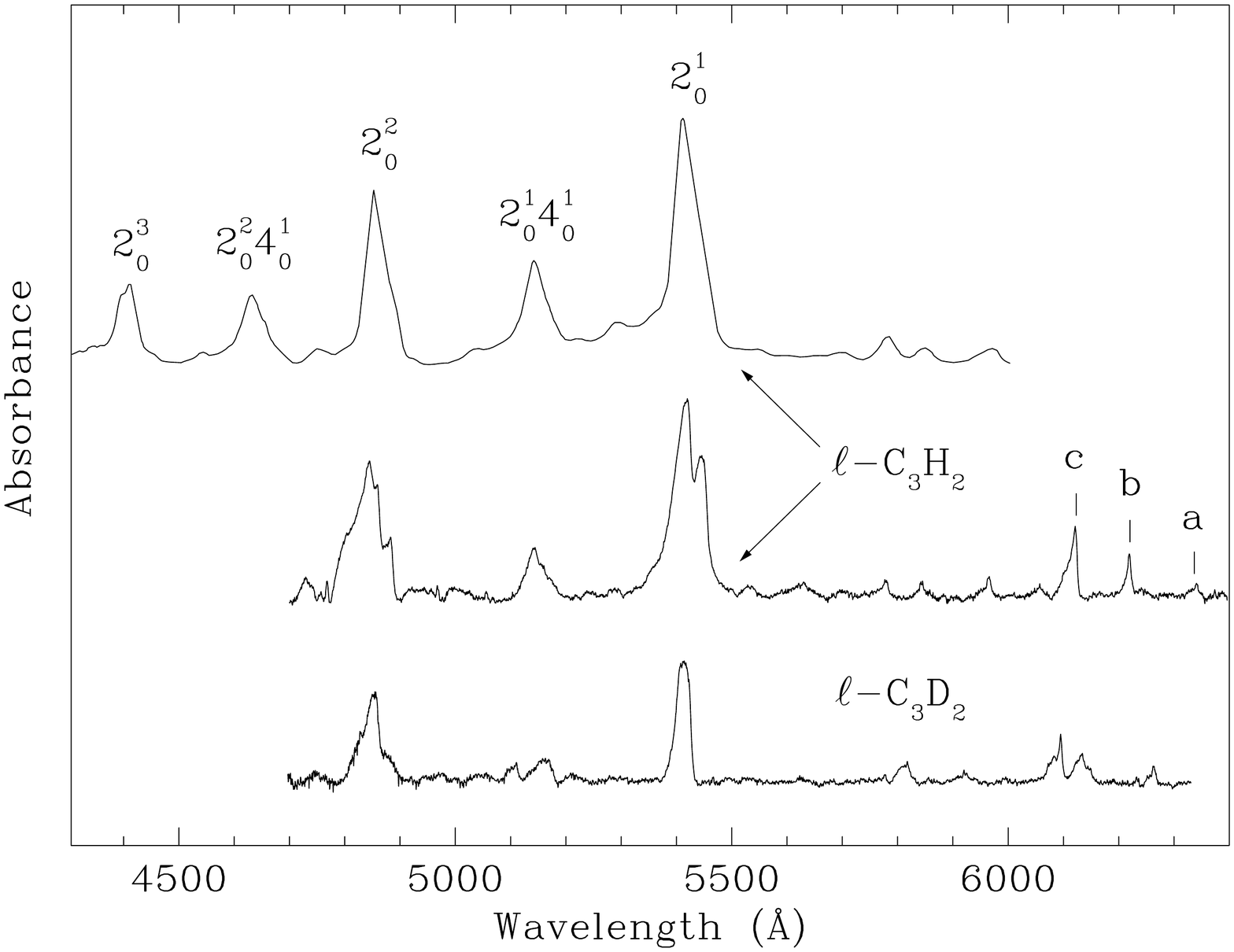}
\caption{Absorption spectra of $l$-C$_3$H$_2$ measured in 6\,K neon matrices. The top trace was obtained by 1470\,\AA\/ photolysis of allene embedded in the matrix. It is identical with the published spectrum in \citep{Hodges:2000} generated by another chemical pathway. The middle and bottom traces were observed using mass-selected deposition as described in the experimental section. Labels a,\,b,\,c refer to the rotationally resolved gas-phase measurements \citep{Achkasova:2006,Birza:2005}.\label{mtx} }
\end{figure}
\clearpage

\begin{figure}
\plotone{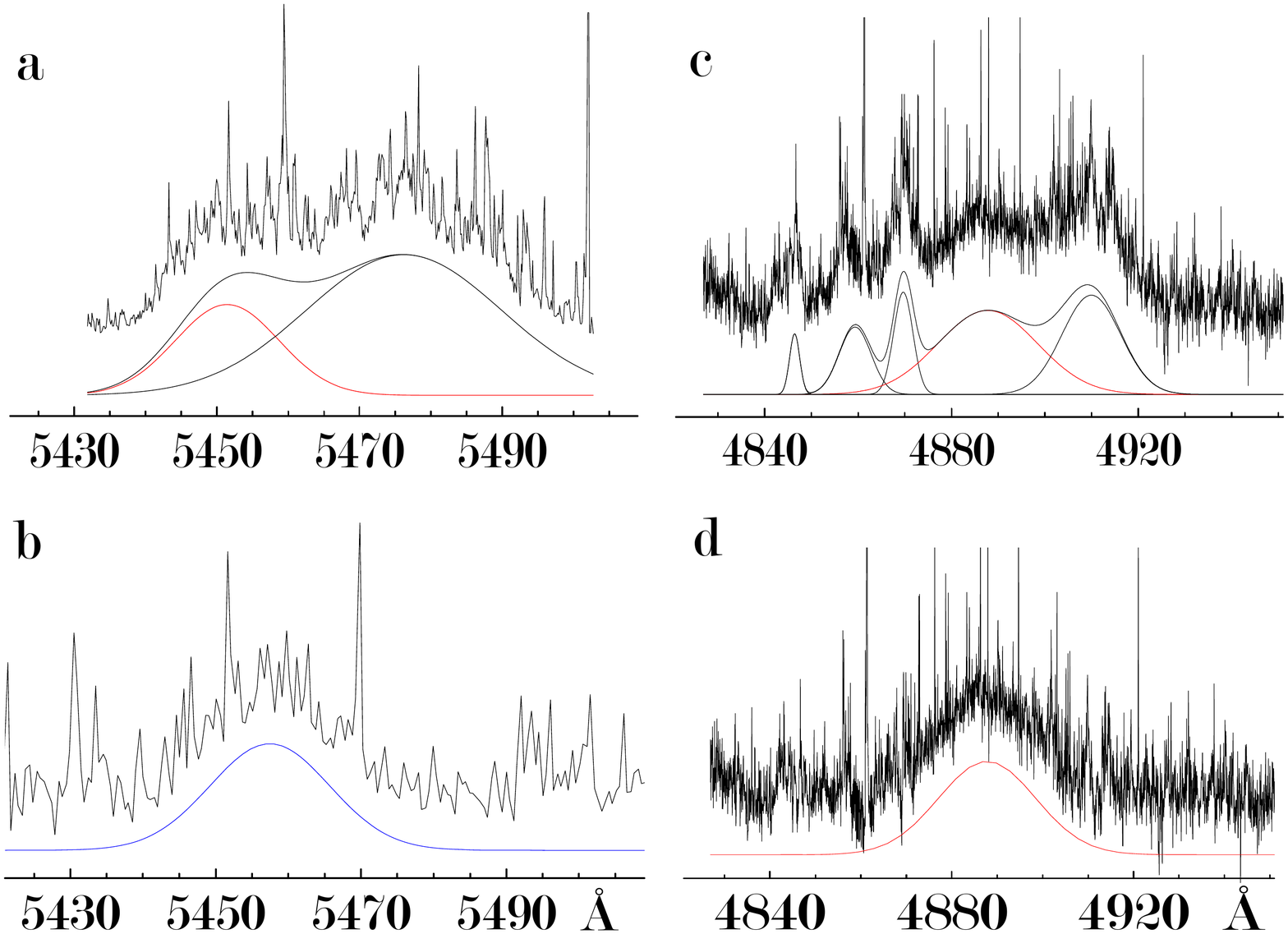}
\caption{Gas-phase laboratory spectra recorded by cavity ring-down spectroscopy in a supersonic slit jet expansion discharge using allene (traces a,\,c,\,d) and DC$_2$D (trace b) as precursors. The continuous, solid lines are Gaussians fitted to the data, after partial removal of the C$_2$ lines and subtraction of the known $l$-C$_3$H absorptions (trace\,c). Absorptions in red are due to $l$-C$_3$H$_2$.\label{cavity} }
\end{figure}
\clearpage

\begin{figure}
\plotone{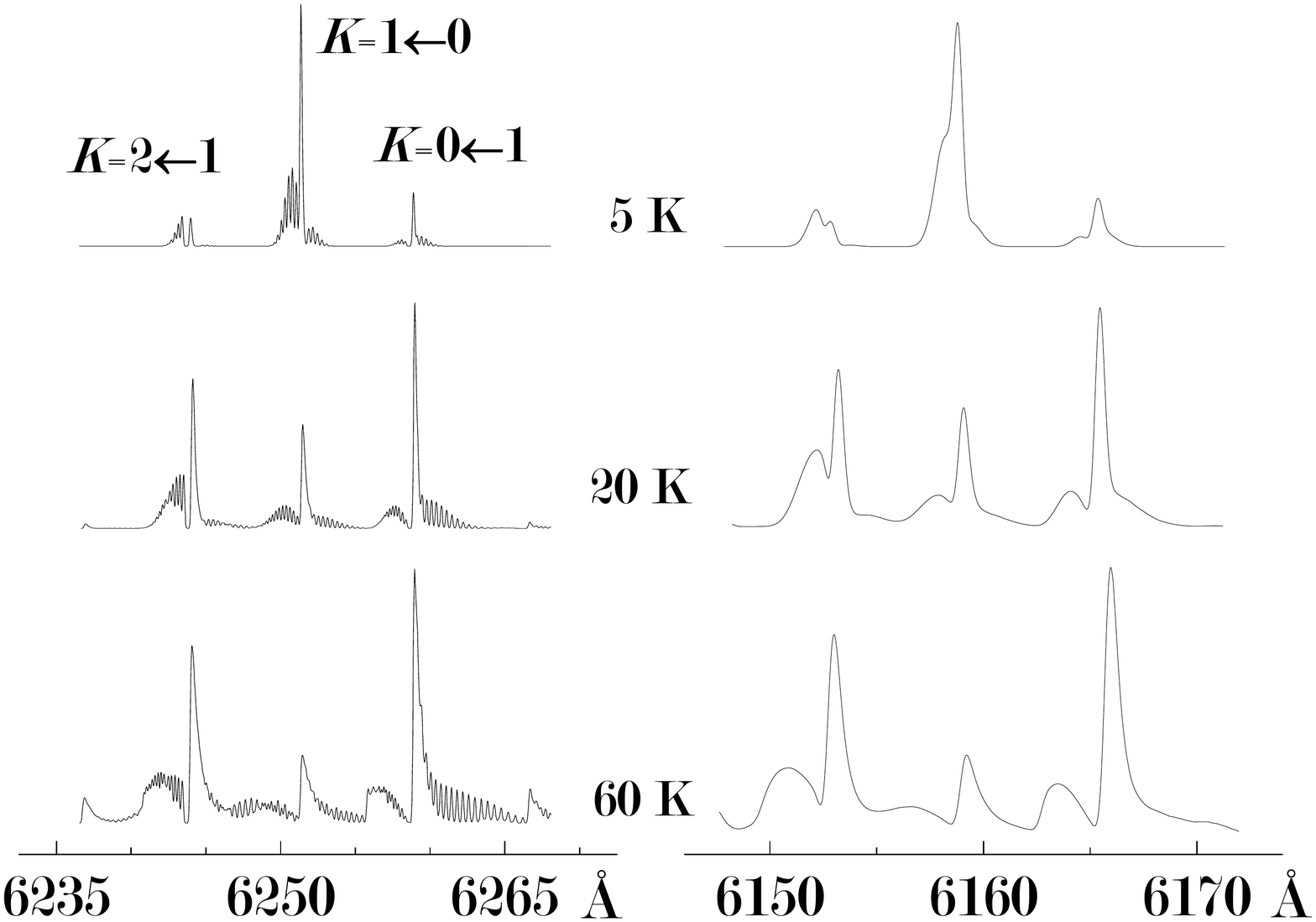}
\caption{Simulated rotational profiles of the b (left) and c bands (right) of $l$-C$_3$H$_2$ which have been observed in the gas phase earlier. The structure is characteristic of an asymmetric top, showing the three $K$ components 2$\leftarrow$1, 1$\leftarrow$0 and 0$\leftarrow$1. The plot on the left uses 0.15\,\AA\/ linewidth for the individual rotational transition, corresponding to 40,000 resolving power used in the astronomical observations. On the right 0.4\,\AA\/ is used, compatible with the conditions of the astronomical observations as the pattern and broadening of the potassium atomic lines show.\label{simT} }
\end{figure}
\clearpage

\begin{figure}
\plotone{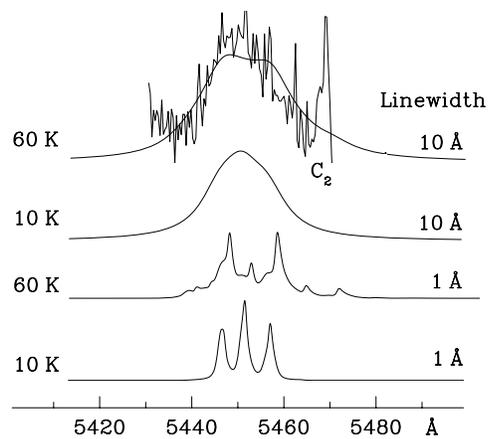}
\caption{Simulation of the 5450\,\AA\/ band profile of $l$-C$_3$H$_2$ $B\,^1$B$_1$\,$\leftarrow$\,$X\,^1$A$_1$ transition at 10 and 60\,K taking 1 or 10\,\AA\/ linewidth. The laboratory absorptions shown in Fig.\,2 were measured with a rotational temperature of around 60\,K, determined by the rotational line intensity of the $l$-C$_3$H$_2$ bands above 6000\,\AA\/. The 10\,\AA\/ linewidth corresponds to the 100\,fs lifetime of the excited electronic state.\label{sim5450} }
\end{figure}
\clearpage

\clearpage

\begin{figure}
 \plotone{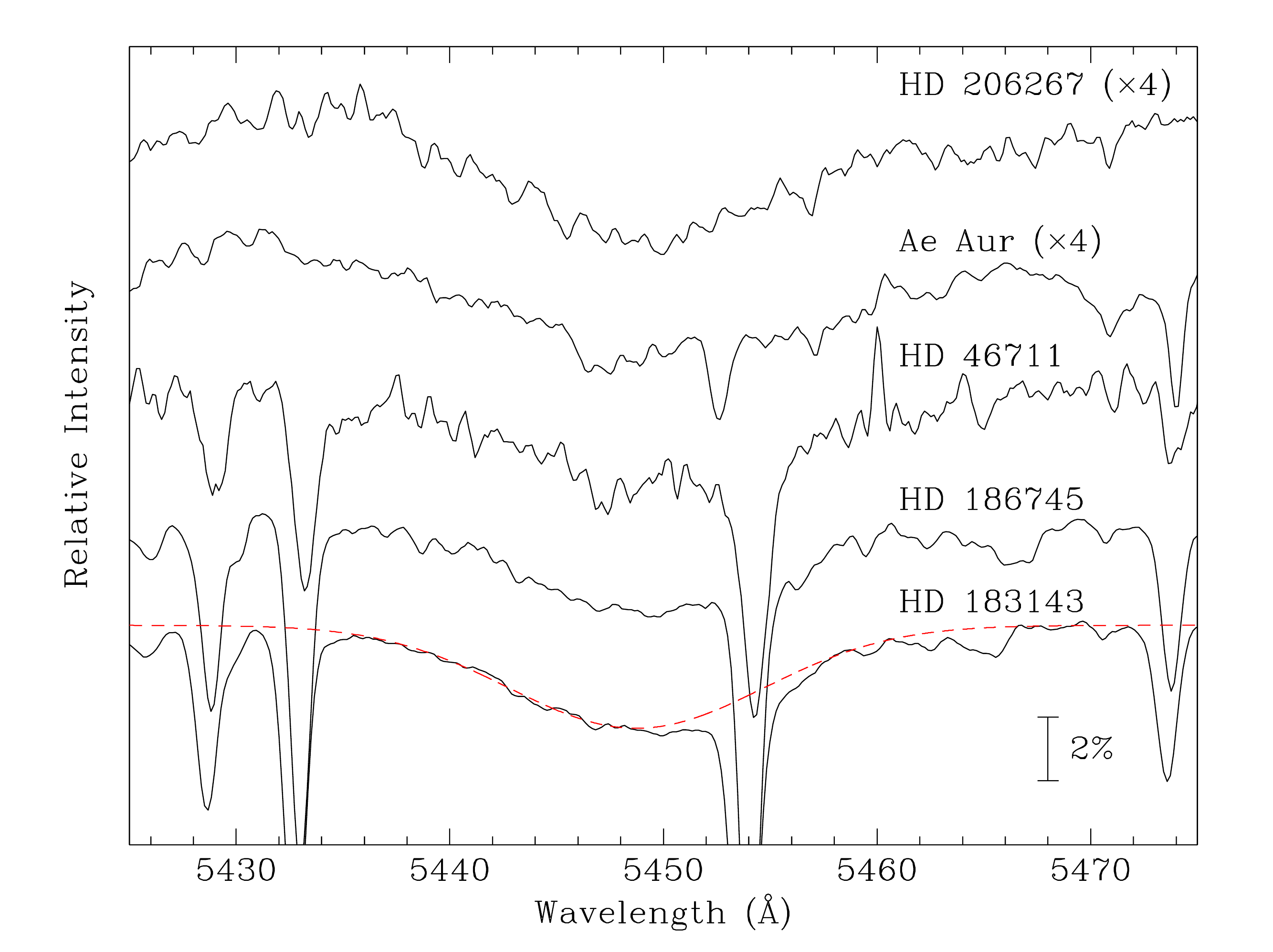}

\caption{Spectra of HD\,46711, 206267, 186745, 183143 and AE\,Aurigae showing the broad DIB at 5450\,\AA\/. The scales for the lightly reddened stars HD\,206267 and AE\,Aurigae have been expanded by a factor 4. The gas-phase lab spectrum of $l$-C$_3$H$_2$ is shown by the dashed, red curve superimposed on the HD\,183143 spectrum. This confirms the agreement found earlier \citep{Linnartz:2010}.\label{5450} }
\end{figure}
\clearpage

\clearpage

\clearpage

\clearpage

\begin{deluxetable}{ccccccccc}
\tablecolumns{9}
\tabletypesize{\footnotesize}
\tablewidth{0pt}
\tablecaption{Absorption band maxima (\AA\/) of $l$-C$_3$H$_2$ and $l$-C$_3$D$_2$ measured in a 6\,K neon matrix and in the gas phase ($K$=1$\leftarrow$0).\label{tab1} }
\tablehead{
\multicolumn{4}{c}{Neon Matrix} & \colhead{} & \multicolumn{2}{c}{Gas Phase} & \colhead{Transition} & \colhead{Band} \\
\cline{1-4} \cline{6-7} \\
\colhead{$\lambda$($l$-C$_3$H$_2$)} &
\colhead{$I$} &
\colhead{$\lambda$($l$-C$_3$D$_2$)} &
\colhead{$I$} &
\colhead{} &
\colhead{$\lambda$($l$-C$_3$H$_2$)} &
\colhead{$\lambda$($l$-C$_3$D$_2$)} &
\colhead{} &
\colhead{}
}
\startdata
6284 & 0.2 & 6260 & 0.4 & & 6318.9 & & $A\,^1$A$_2$\,$\leftarrow$\,$X\,^1$A$_1$ & a \\
6219 & 0.3 & 6130 & 1.4 & & 6251.7 & & $''$ & b \\
6122 & 0.9 & 6093 & 1.5 & & 6159.2 & & $''$ & c \\
5445/5417\tablenotemark{1} & 10 & 5412  & 10 & & 5450(3) & 5458(3) & $B\,^1$B$_1$\,$\leftarrow$\,$X\,^1$A$_1$ & 2$_0^1$ \\
5143 & 3 & 5160 & 3.5 & & 5165$-$5185\tablenotemark{2} & & $''$ & 2$_0^1$\,4$_0^1$ \\
4856 & 9 & 4857 & 12 & & 4887(3) & & $''$ & 2$_0^2$ \\
4633 & 4.4 & & & & 4645$-$4665\tablenotemark{2} & & $''$ & 2$_0^2$\,4$_0^1$ \\
4412 & 4.7 & & & & 4425$-$4445\tablenotemark{2} & & $''$ & 2$_0^3$
\enddata
\tablenotetext{1}{most intense site}
\tablenotetext{2}{extrapolated values based on neon/gas shifts of the 2$_0^1$ band}
\end{deluxetable}
\clearpage

\begin{deluxetable}{ccccrccc}
\tablecolumns{8}
\tabletypesize{\footnotesize}
\tablecaption{4881 and 5450\,\AA\/ DIB central depths (\%) and EW (m\AA\/).\label{tab2} }
\tablewidth{0pt}
\tablehead{
\colhead{Star} &
\colhead{Sp} &
\colhead{$E$($B-V$)} &
\multicolumn{2}{c}{4881} &&
\multicolumn{2}{c}{5450} \\
\cline{4-5} \cline{7-8} \\
\colhead{} &
\colhead{} &
\colhead{} &
\colhead{\%} &
\colhead{EW\tablenotemark{1}} &&
\colhead{\%} &
\colhead{EW\tablenotemark{2}} }
\startdata
HD\,46711& B3 II &1.05& 5.7&1425&&3.0&390\\
183143& B7 Ia& 1.28& 6.6&1650&&3.0&390\\
186745&B8 Ia&0.78&&&&2.6&338\\
190603&B1.5 Ia+& 0.70&4.3&1075&&&\\
AE\,Aur &O9 V&0.48&&&&0.8&104\\
206267&O6&0.53&&&&1.0&130\\
\enddata
\tablenotetext{1}{based on the measured \% central depth and a FWHM of 25\,\AA}
\tablenotetext{2}{based on the measured \% central depth and a FWHM of 13\,\AA}
\end{deluxetable}
\clearpage

\begin{deluxetable}{ccclcc}
\tablecolumns{6}
\tabletypesize{\footnotesize}
\tablecaption{6000\,\AA\/ triplet targets.\label{tab3} }
\tablewidth{0pt}
\tablehead{
\colhead{Star} &
\colhead{Sp} &
\colhead{$E$($B-V$)} &
\colhead{IS K\,I\tablenotemark{1}} &
\multicolumn{2}{c}{Spectrograph} \\
\cline{5-6} \\
&&&km\,s$^{-1}$&\colhead{6150}&\colhead{6250} }
\startdata
$\zeta$\,Oph& O9.5 V &0.30&$-$14.53 & Esp\tablenotemark{2}&Gec\tablenotemark{3}\\
HD\,179406& B3 V&0.27&$-$12.53& & Esp  \\
204827&B0 V& 1.11& $-$17.0 &Esp&Esp\\
206267&O6 e&0.53& $-$17.0 &&Gec\\
\enddata
\tablenotetext{1}{for strongest component---see Fig.\,8}
\tablenotetext{2}{ESPaDONs $R$=80,000}
\tablenotetext{3}{Gecko $R$=120,000}
\end{deluxetable}
\clearpage

\begin{deluxetable}{ccccccccccccc}
\tabletypesize{\footnotesize}
\tablecaption{Comparison of laboratory and interstellar 6250\,\AA\/ triplet component wavelengths $\lambda$ (\AA\/)\tablenotemark{1} and EW (m\AA\/). The lab values give the maximum of the $K$=1$\leftarrow$0 band using a Lorentzian width of $\sim$0.15\,\AA\/.\label{tab5} }
\tablewidth{0pt}
\tablehead{
\colhead{Source}&
\colhead{$\lambda$}&
\colhead{}&
\colhead{$\lambda$}&
\colhead{EW}&
\colhead{$\lambda$}&
\colhead{EW}&
\colhead{$\lambda$}&
\colhead{EW}&
\colhead{$\lambda$}&
\colhead{EW}&
\colhead{$\lambda$}&
\colhead{EW}}
\startdata
lab&6152.8 &&6159.0&&6165.8&&6244.4&&6251.6&&6259.0&\\ 
\\
HD\,179406 &  &&&&&&6244.4& 2.4&6251.0&1.6&6259.5&2.2\\
                    &&&&&&  &6245.3&1.5&6252.5&1.0&&\\
\\
HD 206267 &  &&&&&&6244.5 & 2.3  &6251.0 & 1.2 &6259.5 & 1.1\\
                    &&&&&& &6245.3 & 1.5& 6252.2& 0.4& \\
\\
HD 204827&&&6158.5&8.0&6165.8&3.0&6244.6 & 5.0 &6251.0&  4.3&6259.4 & 3.0\\
                 &&&&&&    &6245.2 & 1.5 & 6252.2 & 2.6&\\
\\
$\zeta$\,Oph&&&6158.5&1.4&6165.8&1.6&6244.5&  0.5& 6251.0 & 0.6&6258.4&  0.2\\
                      &&&&&&&6245.3 & 0.3& 6252.2&  0.3 &6259.6 & 0.5\\
\enddata
\tablenotetext{1}{some interstellar components are paired}
\end{deluxetable}
\clearpage


\begin{thebibliography}{}
\bibitem[Achkasova et al.(2006)]{Achkasova:2006} Achkasova,~E., Araki,~M., Denisov,~A. and Maier,~J.~P. 2006, J. Mol. Spectrosc., 237, 70
\bibitem[Biennier et al.(2003)]{Biennier:2003} Biennier,~L., Salama,~F., Allamandola,~L.~J. and Scherer,~J.~J. 2003, J. Chem. Phys., 118, 7863
\bibitem[Birza et al.(2005)]{Birza:2005} Birza,~P., Chirakolava,~A., Araki,~M., Kolek,~P. and Maier,~J.~P. 2005, J. Mol. Spectrosc., 229, 276
\bibitem[Cernicharo et al.(1991)]{Cernicharo:1991} Cernicharo,~J., Gottlieb,~C.~A., Gu\'elin,~M., Killian,~T.~C., Paubert,~G., Thaddeus,~P. and Vrtilek,~J. 1991, \apj, 368, L39
\bibitem[Cernicharo et al.(1999)]{Cernicharo:1999} Cernicharo,~J., Cox,~P., Foss\'e,~D. and G\"usten,~R. 1999, \aap, 351, 341
\bibitem[Cox et al.(1988)]{Cox:1988} Cox,~P., G\"usten,~R. and Henkel,~C. 1988, \aap, 206, 108
\bibitem[Ding et al.(2001)]{Ding:2001} Ding,~H., Pino,~T., G\"uthe,~F. and Maier,~J.~P. 2001, J. Chem. Phys., 115, 6913
\bibitem[Freivogel et al.(1994)]{Freivogel:1994} Freivogel,~P., Fulara,~J., Lessen,~D., Forney,~D. and Maier,~J.~P. 1994, Chem. Phys., 189, 335
\bibitem[Guethe et al.(2001)]{Guethe:2001} Guethe,~F., Tulej,~M., Pachkov,~M.~V. and Maier,~J.~P. 2001, \apj, 555, 466
\bibitem[Herbig(1995)]{Herbig:1995} Herbig,~G.~H. 1995, \araa, 33, 19
\bibitem[Herbig(2000)]{Herbig:2000} Herbig,~G.~H. 2000, \apj, 542, 334 and references therein
\bibitem[Hobbs et al.(2008)]{Hobbs:2008} Hobbs,~L.~M., York,~D.~G., Snow,~T.~P., Oka,~T., Thorburn,~J.~A., Bishof,~M., Friedman,~S.~D., McCall,~B.~J., Rachford,~B., Sonnentrucker,~P. and Welty,~D.~E. 2008, \apj, 680, 1256
\bibitem[Hodges et al.(2000)]{Hodges:2000} Hodges,~J.~A., McMahon,~R.~J., Sattelmeyer,~K.~W. and Stanton,~J.~F. 2000, \apj, 544, 838
\bibitem[Iglesias-Groth et al.(2008)]{Iglesias:2008} Iglesias-Groth,~S., Manchado,~A., Garc\'ia-Hern\'andez,~D.~A., Gonz\'alez Hern\'andez,~J.~I. and Lambert,~D.~L. 2008, \apjl, 685, L55
\bibitem[Jacox(1994)]{Jacox:1994} Jacox,~M.~E. 1994, J. Phys. Chem. Ref. Data, 3, 1
\bibitem[Jenniskens \& D\'esert(1994)]{Jenniskens:1994} Jenniskens,~P. and D\'esert,~F.$-$X. 1994, \aaps, 106, 39
\bibitem[Kre{\l}owski et al.(2010)]{Krelowski:2010} Kre{\l}owski,~J., Beletsky,~Y., Galazutdinov,~G.~A., Ko{\l}os,~R., Gronowski,~M. and LoCurto,~G. 2010, \apjl, 714, L64
\bibitem[Linnartz et al.(1998)]{Linnartz:1998} Linnartz,~H., Motylewski,~T. and Maier,~J.~P. 1998, J. Chem. Phys., 109, 3819
\bibitem[Linnartz et al.(2010)]{Linnartz:2010} Linnartz,~H., Wehres,~N., Van Winckel,~H., Walker,~G.~A.~H., Bohlender,~D.~A., Tielens,~A.~G.~G.~M., Motylewski,~T. and Maier,~J.~P. 2010, \aap, 511, L3
\bibitem[Liszt \& Lucas(2000)]{Liszt:2000} Liszt,~R. and Lucas,~R. 2000, \aap, 358, 1069
\bibitem[Liszt et al.(2006)]{Liszt:2006} Liszt,~H.~S., Lucas,~R. and Pety,~J. 2006, \aap, 448, 253
\bibitem[Madden et al.(1989)]{Madden:1989} Madden,~S.~C., Irvine,~W.~M., Mattheus,~H.~E., Friberg,~P. and Swade,~D.~A. 1989, \apj, 97, 1403
\bibitem[Maier et al.(2004)]{Maier:2004} Maier,~J.~P., Walker,~G.~A.~H. and Bohlender,~D.~A. 2004, \apj, 602, 286
\bibitem[McCall et al.(2002)]{McCall:2002} McCall,~B.~J., Hinkle,~K.~H., Geballe,~T.~R., Moriarty-Chieven,~G.~H., Evans,~N.~J., Kawaguchi,~K., Tanako,~S., Smith,~V.~V. and Oka,~T. 2002, \apj, 567, 391
\bibitem[Mebel et al.(1998)]{Mebel:1998} Mebel,~A.~M., Jackson,~W.~M., Chang,~A.~H.~H. and Lin,~S.~H. 1998, J. Am. Chem. Soc., 120, 5751
\bibitem[Motylewski et al.(2000)]{Motylewski:2000} Motylewski,~T., Linnartz,~H., Vaizert,~O., Maier,~J.~P., Galazutdinov,~G.~A., Musaev,~F.~A., Kre{\l}owski,~J., Walker,~G.~A.~H. and Bohlender,~D.~A. 2000, \apj, 531, 312
\bibitem[Noller et al.(2009)]{Noller:2009} Noller,~B., Margraf,~M., Schröter,~C., Schultz,~T. and Fischer,~I. 2009, Phys. Chem. Chem. Phys., 11, 5353
\bibitem[Oka et al.(2003)]{Oka:2003} Oka,~T., Thorburn,~J.~A., McCall,~B.~J., Friedman,~S.~D., Hobbs,~L.~M., Sonnentrucker,~P., Welty,~D.~E. and York,~D.~G. 2003, \apj, 582, 823
\bibitem[Pan et al.(2004)]{Pan:2004} Pan,~K., Federman,~S.~R., Cunha,~K., Smith,~V.~V. and Welthy,~D.~E. 2004, \apjs, 151, 313
\bibitem[Seburg et al.(1997)]{Seburg:1997} Seburg,~R.~A., Patterson,~E.~V., Stanton,~J.~F. and McMahon,~R.~J. 1997, J. Am. Chem. Soc., 119, 5847
\bibitem[Snow \& McCall(2006)]{Snow:2006} Snow,~T.~P. and McCall,~B.~J. 2006, \araa, 44, 367
\bibitem[Thorburn et al.(2003)]{Thorburn:2003} Thorburn,~J.~A., Hobbs,~L.~M., McCall,~B.~J., Oka,~T., Welty,~D.~E., Friedman,~S.~D., Snow,~T.~P., Sonnentrucker,~P. and York,~D.~G. 2003, \apj, 584, 339
\bibitem[Vrtilek et al.(1990)]{Vrtilek:1990} Vrtilek,~J.~M., Gottlieb,~C.~A., Gottlieb,~E.~W., Killian,~T.~C. and Thaddeus,~P.~A. 1990, \apj, 364, L53
\end{thebibliography}
\end{document}